\documentclass[10pt,showkeys,showpacs,amsmath,amssymb,prl]{revtex4}
\usepackage[english]{babel}
\usepackage[T2A]{fontenc}
\usepackage[cp866]{inputenc}
\usepackage[dvips]{graphicx}
\usepackage{amsmath,euscript,amssymb}
\newcommand{\be}{\begin{equation}}
\newcommand{\ee}{\end{equation}}
\newcommand{\bea}{\begin{eqnarray}}
\newcommand{\eea}{\end{eqnarray}}
\newcommand{\dd}{\mathrm{d}}
\def\gsim{\mathrel{\raise.3ex\hbox{$>$\kern-.75em\lower1ex\hbox{$\sim$}}}}
\def\lsim{\mathrel{\raise.3ex\hbox{$<$\kern-.75em\lower1ex\hbox{$\sim$}}}}

\begin{document}
\title{Higgs Particle Mass in Cosmology}
\author{A.B. Arbuzov}
\author{L.A. Glinka}
\author{V.N. Pervushin}
\affiliation{Joint Institute for Nuclear Research, 141980 Dubna,
Russia}
\date{\today}



\begin{abstract}
 A version of the Standard Model is considered, where the electroweak
symmetry breaking is provided by  cosmological initial data given
for the zeroth Fourier harmonic of the Higgs field $\langle\phi\rangle$.
The  initial data symmetry breaking mechanism removes
the Higgs field contribution to the vacuum energy density,
possible creation of monopoles, and tachion behavior at high energies,
if one imposes  an ``inertial'' condition on the Higgs potential
$\textsf{V}_{\rm Higgs}(\langle\phi\rangle)=0$.
The requirement of zero radiative corrections to this {\em inertial} condition coincides
with the limiting point of the vacuum stability in the Standard Model.
The latter together with the direct experimental limit gives the prediction
for the mass of the Higgs boson to be in the range $114 < m_h \lsim 134$~GeV.
\end{abstract}

\maketitle

\section{Introduction} \label{sect1}

The discovery and study of the Higgs boson  are of the
highest priority for the modern elementary particle physics~\cite{higgs,hhg}.
The accepted description of the Higgs field is based on the  classical Higgs
potential.

However, there is a well know list of consequences (including the
tremendous potential vacuum energy density,  possible creation of monopoles, a
tachion behavior at high energies, a fine tuning required to avoid
the triviality and instability bounds,
and so on) that  are
incompatible with cosmological observations \cite{Zeldovich_78,Linde_90,1_06,2_06}.

In the present paper, we suggest  to overcome  these problems,
by considering a model, with a special condition on the Higgs potential in a single
point, which provides that the Higgs field contribution to the vacuum energy
density is zero. The very statement of the problem assumes
that  the condition should be established within Cosmology,
 and the zeroth harmonic of the Higgs field should have a dynamical status \cite{linde1}.
Introduction of a condition on the potential can be unambiguously performed
if we have nontrivial initial data in the dynamical equations.
For this reason we start with a derivation of cosmological equations
in the framework of the Hilbert variation principle with constraints of initial data.

The paper is organized as follows. First we formulate a cosmological model
separating zeroth harmonics of  all scalar fields in the General Relativity (GR) and
the Standard Model (SM).
In Sect.~3 the zero mode initial data problem is discussed on the classical level.
The SM particle contributions into the cosmological energy density  are considered in Sect.~4
on the quantum level.
The Higgs effect in the cosmological model is studied in Sect.~5.
A discussion of results is given in Conclusion. Through out the paper we will use the units
 \be\label{unit-1}
 \hbar=c=M_{\rm Planck}\sqrt{\dfrac{3}{8\pi}}=1.
 \ee

\section{The cosmological approximation } \label{sect2}

Let us start with the General Relativity given by the sum of
 Hilbert's  action~\cite{H} and  the SM one~\cite{weak} supplemented by an additional scalar field $Q$
 governing the Universe evolution \cite{Guth_92}
 \be\label{1-1}
 S_{\rm GR}=\int\limits_{ }^{ }d^4x\sqrt{-g}\left[-\frac{1}{6}R(g)
 +{\partial_\mu\phi\partial^\mu\phi}
 +{\cal L}_{\rm SM}(\phi)+\partial_\mu Q\partial^\mu Q -\textsf{V}_{\rm U}(Q)\right].
 \ee
The Riemannian space-time with the interval $ds^2=g_{\mu\nu}dx^\mu dx^\nu$ is assumed.
The Standard Model Lagrangian depends on the Higgs field $\phi$ in the usual way:
 \bea \label{sm-2}
 {\cal L}_{\rm SM}(\phi)&=& -\phi\sum_f g_f\bar f
 f+\frac{{\phi^2}}{2}\sum_{ v} g^2_{ v}v_\mu v^\mu -\textsf{V}_{\rm Higgs}(\phi)
 + {\cal L}_{\rm SM}(\phi=0).
 \eea
Here we separated terms with Higgs coupled to vector $(v)$ and  fermion $(f)$ fields,
and the  potentials of  scalar fields $\phi,Q$.

  Modern cosmological models \cite{Wein_73,1_06,2_06,MFB,fuk,WMAP01}
 are based on the so-called cosmological principle introduced by Einstein
 \cite{Einstein_17}.
 In his model, matter is
evenly distributed in the Universe and the cosmological
time is defined so that local characteristics of the Universe
averaged over a large enough area depend only on
this time \cite{fried_22}.
Recall that, in  the modern  models,  local scalar characteristics of the Universe evolution
 averaged over a large coordinate volume $V_0=\int d^3x $ (\textit{i.e.} zeroth harmonics)
\bea\label{z-s1-1}
 \log a\equiv\dfrac{1}{6V_0
 }\int d^3x \log |g^{(3)}|,~~~~\langle
 \phi\rangle\equiv\dfrac{1}{V_0
 }\int d^3x \phi,~~~~~\langle
 Q\rangle\equiv\dfrac{1}{V_0
 }\int d^3x \,Q
 \eea
depend only on the cosmological time $dt =a(\eta)d\eta$
 of the conformal-flat  interval
 \be \label{cti-2}
 ds^2=a^2(\eta)[(d\eta)^2-(dx^j)^2],
 \ee
 where $d\eta=N_0(x^0)dx^0$ is the conformal time of a photon on its light cone $ds^2=0$, and
$N_0(x^0)=\langle \sqrt{-\widetilde{g}}\,\widetilde{g}^{00}\rangle^{-1}$
 is the global lapse function arising in  the second term of action
 \be\label{eta-1}
 S_{\rm GR}[g=a^2\widetilde{g},f=a^{-3/2}\widetilde{f},\phi=\widetilde{\phi}a^{-1}]\equiv
 S_{\rm GR}[\widetilde{g},\widetilde{f},\widetilde{\phi}]
 +V_0\int\limits_{\eta=0}^{\eta_0} d\eta aa'', \qquad a'=da/d\eta
 \ee
after the conformal transformations of  fields in  action (\ref{1-1}) \cite{Barbashov_06,Glinka_08}.
Then the separation of the zeroth harmonics
 \be\label{dec-1}\phi=\langle\phi\rangle+{h}/{\sqrt{2}},~~~~~~
 Q=\langle Q\rangle+{q}/{\sqrt{2}}
 \ee
from the nonzero ones~$\int \dd^3x \, {h}=0$
 associated with scalar particles
 determines a cosmological model  in a flat space-time $\widetilde{ds}^2=(d\eta)^2-(dx^j)^2$.
  Following   \cite{WDW,Misner_69} we shall consider
  this conformal-flat  cosmological approximation of
 the Hilbert action (\ref{1-1}) in the Dirac Hamiltonian approach \cite{dir}
 \bea
\label{4-s}
 S &=& \int\limits_{}^{}\! dx^0
 \!\int d^3x \sum\limits_{F=f,h,q,v}^{} P_{\widetilde{F}}\partial_0\widetilde{F} \!+\int\!\!
 \left\{\!P_{\langle Q\rangle}d\langle Q\rangle\!+\!P_{\langle \phi \rangle} d\langle \phi \rangle
\! -\!P_{\log a}d\log a\!+\! \textsf{C}_{\rm U}[P,\widetilde{F}]\frac{N_0(x^0)}{4V_0a^2}dx^0\right\},
 \eea
 where $P_{\widetilde{F}}$,
 $P_{\log a}=2V_0a a'$,
$P_{\langle \phi \rangle}=2a^2V_0\langle\phi\rangle'$, and
$ P_{\langle Q \rangle}=2a^2V_0\langle Q\rangle'
  $
  are  canonical conjugate momenta. The global lapse function $N_0$ is the Lagrange multiplier
  so that the
  variation of action (\ref{4-s}) with respect to this lapse function,
  $\dfrac{\delta S}{\delta N_0}=0$, leads to the energy constraint
    \be\label{m-0}
  \textsf{C}_{\rm U}[P,\widetilde{F}]\equiv P_{\log  a}^2-\textsf{E}_{\rm U}^2(a)=0.
  \ee
The quantity
 \be\label{scm-4e}
  \textsf{E}_{\rm U}^2(a)\equiv P_{\langle \phi \rangle}^2+P_{\langle Q\rangle}^2+4V_0^2
 a^6\left[\textsf{V}_{\rm Higgs}(\langle\phi\rangle)+\textsf{V}_U(\langle Q\rangle)\right]+
 4V_0a^2 \textsf{H}(a\langle\phi\rangle|\widetilde{F}),
 \ee
can be  considered as the square of the Universe
 energy, because $\log a$  is treated as
 the Universe evolution parameter in the Wheeler-DeWitt
 field space of events $[\log a|\langle\phi\rangle, \langle Q\rangle,\widetilde{F}]$ \cite{WDW,Misner_69}, and $\textsf{H}(a\langle\phi\rangle|\widetilde{F})$
 is the Hamiltonian of the SM with  masses scaled by the scale factor $m=m_{F0}a(\eta)$.
 Recall that in the case of the Higgs potential $\textsf{V}_{\rm Higgs}=\lambda (\langle\phi\rangle^2-c_0^2)^2$ the masses of vector ($Z,W$), fermion ($f$), and  Higgs ($h$) particles:
 \bea
\label{0W-1}
M_{W}&=&\langle\phi\rangle g_W,
\qquad
M_{Z} = \langle\phi\rangle\sqrt{g^2+g'^2},
\qquad
m_{f}= \langle\phi\rangle g_f,
\qquad
m_h = [4\lambda\langle\phi\rangle^2+2(\langle\phi\rangle^2-c_0^2)]^{1/2}
 \eea
arise in the lowest order in the coupling constant.
Quantity $\langle\phi\rangle$ is the solution of the equations of motion
 following from
the  emerging cosmological GR\&SM action (\ref{4-s}).

\section{Initial Data and Observable Variables in Cosmology\label{sect3}}

It is reasonable
to define initial data  in terms of conformal time, because the
coordinate-distance --- redshift relation $r(z)$
is determined by the constraint
$P_{\log a}=\pm \textsf{E}_{\rm U}(a)=2V_0aa'$ and the light-cone interval $\widetilde{ds}^2=d\eta^2-dr^2=0$,
so that
 \bea \label{cit-4c}
 dr(z)= d\eta=\pm 2V_0 \frac{a da}{\textsf{E}_{\rm U}(a)}\Big |_{\,a=(z+1)^{-1}}.
 \eea
Therefore, we look at the initial data problem by analyzing the constraint-shell value of
the action  (\ref{4-s}):
 \bea \label{5-s}
 S^{(\pm)}\Big |_{\, \textsf{C}_{\rm U}[P,\widetilde{F}]=0}=\int\limits_{a_I}^{a_0}d\log a\left\{
 \left[\int d^3x \sum_{\widetilde{F}} P_{\widetilde{F}}\partial_{\log a}\widetilde{F}\right]+
 P_{\langle Q \rangle}\partial_{\log a}\langle Q\rangle+P_{\langle \phi \rangle} \partial_{\log a}\langle \phi \rangle
] \mp \textsf{E}_{\rm U}(a)\right\},
\eea
where the role of the evolution parameter is played by the logarithm of the cosmological
scale factor. It is  accepted \cite{Linde_90} that
 the initial instance $\eta=0$ is absolute,
 there  is the time arrow $\eta \geq 0$, and the primordial value of
 the scale factor was very small. In particular, the Inflationary model~\cite{Linde_90}
 assumes  the Planck epoch, where $a(\eta=0)=a_I\sim 10^{-61}$ in units~(\ref{unit-1}).
 Following the Planck epoch hypothesis,
 we assume that at the initial instance $\eta=0$
 there can be nontrivial  data  for
 the zeroth harmonics  (\ref {z-s1-1}):
 \bea\label{in-d1}
 && a(\eta=0)=a_I, \qquad
 P_{\log a_I}=\textsf{E}_{\rm U}(a_I),
  \\\label{in-d2}
  &&\langle\phi\rangle(\eta=0)=\phi_I, \qquad
 P_{\langle \phi\rangle_I}=2V_0H_\phi,\\\label{in-d3}
&& \langle Q\rangle(\eta=0)=Q_I, \qquad
 P_{\langle Q\rangle_I}=2V_0H_Q;
 \eea
whereas all initial data for local fields are equal zero,
{\it i.e.} there were no any particle-like excitations.
Therefore, at the Planck epoch,
one can neglect contributions of all fields
except the ones of the scalar field zeroth modes. Note also that for the Planck epoch value $a_I\sim 10^{-61}$
the contribution
to the cosmological equation~(\ref{scm-4e}) of the scalar field potentials
$ a^6\left[\textsf{V}_{\rm Higgs}(\langle\phi\rangle)+\textsf{V}_U(\langle Q\rangle)\right]$
is suppressed by the factor $a^6\sim 10^{-366}$ in comparison with the kinetic energy.
On the classical level, the Universe energy ~(\ref{scm-4e}) in the neighborhood of
the cosmological singularity point, $a=0$, takes the form
\be\label{a=0}
 \textsf{E}_{\rm U}(a\ll 1) \simeq \textsf{E}_{\rm U}(0)+2V_0a^2\frac{ \textsf{H}(0|\widetilde{F})}{\textsf{E}_{\rm U}(0)}\, ,
 \ee
where
\bea\label{a=0-1}
\textsf{E}_{\rm U}(0) \equiv \sqrt{P_{\langle \phi \rangle}^2+P_{\langle Q\rangle}^2}=2V_0\sqrt{H_\phi^2+H_Q^2}
=2V_0 H_0\Omega^{1/2}_{\rm rigid}
\eea
is the potential-free energy of inertial motion of the zeroth scalar field harmonics.
The field Hamiltonian $\textsf{H}(0|\widetilde{F})$ in this limit looks like the one of the massless Standard Model in the flat
space-time with interval $\widetilde{ds}^2=(d\eta)^2 - (dx^k)^2$
and the conformal time  (\ref{cit-4c})
\bea\label{a=0-3}
d\eta = 2V_0\frac{a d a}{\textsf{E}_{\rm U}(0)}=\frac{a d a}{H_0{\Omega^{1/2}_{\rm rigid}}}\, .
\eea
Due to~(\ref{a=0}) and~(\ref{a=0-3}) the constraint-shell action~(\ref{5-s}) is a sum of the cosmological and field actions:
\bea \label{6-s}
S^{(\pm)}(1\gg a \geq a_I) &=& S^{(\pm)}_{\rm rigid} + S^{(\pm)}_{\rm radiation},
\\ \label{7-s}
S^{(\pm)}_{\rm rigid} &=& \int\limits_{\log a_I}^{\log a}d\log \widetilde{a}\left\{
 P_{\langle Q \rangle}\partial_{\log \widetilde{a}}\langle Q\rangle+P_{\langle \phi \rangle} \partial_{\log \widetilde{a}}\langle \phi \rangle
 \mp \textsf{E}_{\rm U}(0)\,\right\},
 \\ \label{8-s}
 S^{(\pm)}_{\rm radiation} &=& \int\limits_{0}^{\eta}d\widetilde{\eta}\left\{
 \left[\int d^3x \sum_{\widetilde{F}} P_{\widetilde{F}}\partial_{\widetilde{\eta}}\widetilde{F}\right]
  \mp \textsf{H}(0|\widetilde{F})\right\}.
\eea
Action~(\ref{7-s}) corresponds to the
most singular primary energetic regime of the Universe rigid state. On the classical level
the particle content of the Universe  described by action~(\ref{8-s})  at  the initial moment
is very poor.

 At the vicinity of $a\to 0$, the considered cosmological model
is reduced to a relativistic conformal mechanics with the constraint on the initial momenta
 \be\label{m-1L}
   \textsf{C}_{\rm U}[P,\widetilde{F}]\equiv P_{\log a}^2-E^2_U(0)=0.
  \ee
  A partial solution of the zero mode equations for the action (\ref{7-s})
\bea \label{L-1}
 \partial_{\log a}P_{\langle \phi \rangle}=0, ~~~~\partial_{\log a}P_{\langle Q \rangle}=0,~~~~
  \partial_{\log a}\langle \phi \rangle=\frac{P_{\langle \phi \rangle}}{\textsf{E}_{\rm U}(0)}, ~~~~ \partial_{\log a}{\langle Q \rangle}=\frac{P_{\langle Q \rangle}}{\textsf{E}_{\rm U}(0)},
 \eea
including the interval (\ref{a=0-3}) takes the form
 \bea
 \label{id-5}
 \langle\phi\rangle(\eta)&=&\phi_I+\frac{P_{\langle \phi \rangle_I}}
 {\textsf{E}_{\rm U}(0)}\log\frac{a(\eta)}{a_I}
 = \phi_I = \frac{M_W}{g_W},\\
 \label{id-6} \langle Q\rangle(\eta)&=&Q_I+\frac{P_{\langle Q \rangle_I}}
 {\textsf{E}_{\rm U}(0)}\log\frac{a(\eta)}{a_I}
 = Q_0+\log{a(\eta)},\\\label{id-4}
 a(\eta)&=&\sqrt{a^2_I+2{\eta}\,H_0\Omega_{\rm rigid}^{1/2}},
 \qquad
\frac{a'}{a}\equiv H(\eta)=\frac{H_0\Omega_{\rm rigid}^{1/2}}{a^2(\eta)}\, .
 \eea
As stated above  the potential terms in the  constraint (\ref{scm-4e})
 are suppressed at the Planck epoch   by the factor $a^6=10^{-366}$
with respect to the contribution of
nonzero initial momenta (\ref{in-d1}) -- (\ref{in-d3}). If the potentials are neglected
in the equations we obtain the  solutions
well known as the rigid state $\Omega_{\rm rigid}\not =0$, when the density is equal to
the pressure. Note that one can assume the trivial initial data for the momentum of
the Higgs field zeroth harmonic:
\bea \label{id-7}
P_{\langle \phi \rangle_I} = 0.
\eea
The averaged value of this harmonic is related to
the Weinberg coupling $g_W$ and the vector boson mass in the standard way (\ref{id-5}).
The initial data for $Q$ field (\ref{id-6}) with nonzero momentum is required to initialize
the Universe evolution in an analogy to {\em inflaton} models.

One can see that the identification of $\log a$ with the evolution parameter
unambiguously determines  the energy in the action (\ref{7-s}) as
       solutions of the energy constraint (\ref{m-1L}) with respect to
  the corresponding canonical momentum $P_{\log a}=\pm \textsf{E}_{\rm U}$ \cite{WDW}.
  Among these solutions there is a negative one. This means that
    the classical system
 is not stable in the field space of events $[\log a|\langle\phi\rangle, \langle Q\rangle]$.  Like a stable orbit of an atomic electron,
the stable  Universe  has a quantum status.
 The primary quantization of the energy constraint  (\ref{m-0}) $\mathcal{C}(P)=0\to \mathcal{C}(\hat P)\Psi=0$
and the secondary one
$ \Psi\rightarrow \hat {\mathbf{\Psi}}=(2\textsf{E}_{\rm U})^{-1/2}[\hat A^++\hat A^-];~~[\hat A^-,\hat A^+]=1$
with the vacuum postulate $\hat A^-|0>=0$ give us the traffic
rules in the field space of events \be \label{1-2v}  P_{\log a}\geq 0, ~~~~a_I < a;~~~~~P_{\log
a}\leq 0, ~~~~a_I > a \ee
 and the arrow of
  time $\eta\geq 0$ is given by Eq. (\ref{cit-4c}) for both values of the energy $P_{\log a}=\pm \textsf{E}_{\rm U}$
  \cite{Pervushin_07}.
Thus, the time arrow problem is solved by both the primary  quantization of the energy constraint (\ref{m-1L}) and the secondary one
  in the spirit of  QFT anomalies
 arising with the construction of vacuum as a state with minimal energy~\cite{Pervushin_07}.
One can say that the arrow of
  time $\eta\geq 0$ is the evidence of the quantum nature of our Universe.

As it was discussed yet by Friedmann more than
80 years ago \cite{fried_22} with a reference to the Weyl idea of the conformal symmetry  \cite{we},
  the Einstein General Relativity  (\ref{eta-1}) admits two types of cosmological variables
  and coordinates
  that  can be identified with observable quantities.
These two types are marked
on the left and right hand sides of (\ref{eta-1}) as $F,ds$  and $\widetilde{F},\widetilde{ds}$.
Now both these variables the standard, $(F,ds)$, and conformal, $(\widetilde{F},\widetilde{ds})$ are well-known in current literature \cite{Narlikar}
as two different types of Cosmology: the Standard Cosmology (SC)
with a hot temperature $T_{SC}=T_0/a(t)$, expanded  distances $R_{SC}=ra(t)$,
and constant masses $m_{SC}=m_0$,
 and the
Conformal Cosmology (CC) with constant conformal temperature $T_{CC}=T_0$,
 coordinate distances $R_{CC}=r$, and running masses $m_{CC}=m_0 a(\eta)$
defined by $\langle\widetilde{\phi}\rangle=a\langle\phi\rangle$, respectively~\cite{Behnke_02,Behnke_04,zakhy}.
Standard variables $R,t$ are used as a mathematical tool to solve
the Schr\"odinger wave equation $\widetilde{\Psi}^{(k)}_A(\eta,r)$ with the running mass and size.
It gives
 equidistant spectrum $-i(d/d\eta)\widetilde{\Psi}^{(n)}_A(\eta,r)=[\alpha^2m_0/(2n^2)]
 \widetilde{\Psi}^{(n)}_A(\eta,r)
 $ for any wave lengths of cosmic photons
 remembering the size of the atom at the moment of their emission \cite{Glinka_08}.

 In the first case (SC) we have the
 temperature history of the Universe; whereas in second case (CC), we have the mass evolution, where
 the constant cold Early Universe
looks like the hot one for any particles because their masses are disappearing.

The best fit to 186  high-redshift Type Ia
supernovae and SN1997ff data \cite{Riess_98,Riess_04} requires
cosmological constants $\Omega_{\Lambda}=0.7$ and $\Omega_{\rm Cold
Dark Matter}=0.3$ in the case of the cosmological evolution of lengths (SC).
In the case of the cosmological evolution of masses (CC)
these data  are consistent with the rigid state regime of inertial motion
$\Omega_{\rm rigid}\approx 0.85\pm 0.10$.
In both the cases the Friedmann  equation takes the same form
\bea \label{free-c1}
  \rho (a)&\equiv&
  H^2_0[\Omega_{\rm rigid}+a^2\Omega_{\rm radiation}+a^3\Omega_{\rm M}
 +a^6\Omega_{\rm \Lambda}] = a'^2,
\eea
where $\rho (a)$ is the conformal density and $H_0$ is the Hubble parameter in units (\ref{unit-1}).
In contrast to the SC, the fit in the CC almost does not depend on the
$\Omega_{\rm Cold Dark Matter}$ value~\cite{Behnke_02,Behnke_04,zakhy}.
 %while it predicts reasonable values
%for the $\Omega_{\mathrm{radiation}}$.

Calculation of the primordial helium abundance \cite{Wein_73,Behnke_04}
takes into account $\Omega_{\mathrm{b}}\simeq 4 \cdot 10^{-2}$,
  weak interactions,
 the Boltzmann factor, (n/p) $ e^{\triangle m/T} \sim 1/6$, where $\triangle m$ is the neutron-proton
mass difference, which is the same for both SC and CC,
$\triangle m_{SC}/T_{SC}=\triangle m_{CC}/T_{CC}=(1+z)^{-1}m_{0}/T_{0}$,
and the square root dependence of the z-factor
 on the measurable time-interval $(1+z)^{-1}\sim \sqrt{\rm t_{\rm
measurable} }$ (see Eq. (\ref{id-4})).

 Thus, in CC the rigid state regime initiated by the
 inertial evolution of the scalar field zeroth modes  without any potentials
 is the dominant regime for all epoch including the vacuum creation
 of particles.

\section{Cosmological creation of SM particles} \label{sect4}

 Recall that in QFT observable {\em particles} are identified   with
    holomorphic representation of the conformal field variables
 \begin{equation}\label{hh-1nt}
\widetilde{F}(\eta,\textbf{x})=\dfrac{1}{V_0}\sum_{\textbf{l},\textbf{l}^2\not
=0}c_F(a,\omega_{F,\textbf{l}})\dfrac{e^{i\textbf{\textbf{k}\textbf{x}}}}{\sqrt{2\omega_{F,\textbf{l}}}}
\left[F^+_\textbf{l}(\eta)+F^-_{-\textbf{l}}(\eta)\right],\qquad
\textbf{k}=\dfrac{2\pi}{V^{1/3}_0}\textbf{l},
\end{equation}
in the flat space-time $\widetilde{ds}^2$. Here $c_F(a,\omega_{F,\textbf{l}})$ is the normalized factor that provides
 the free particle Hamiltonian
\be\label{chh-2e}
 \textsf{H}_{\rm free}(am_{F0}|\widetilde{F})
 =\sum\limits_{F, \textbf{l},\textbf{l}^2\not
=0}^{}\biggl[n_{F,\textbf{l}}
+\frac{A_{F}}{2}\biggr]\omega_{F,\textbf{l}}(a)
 \ee
in the  form of  the sum over momenta of
products of occupation numbers
${n_{F,\textbf{l}}}=F^+_\textbf{l} F^-_\textbf{-l}$ and  the one-particle energies $\omega_{F,\textbf{l}}(a)=\sqrt{k^2_{\textbf{l}}+m_{F0}^2a^2}$ ~\cite{grib80,ps1,JINR-1}.
  The zeroth harmonic $\textbf{l}^2 =0$ in the sum (\ref{chh-2e}) is excluded
because the
transverse ($T$) vector and tensor fields  are constructed by means of
the inverse Beltrami-Laplace operator  acting in the
class of functions of nonzero harmonics with the
constraint $\int F d^3x=0$.
 The free particle  Hamiltonian contains %and
   the Casimir energies  \cite{Casimir_48}, positive for
  bosons $A_F=+1$ and negative for fermions $A_F=-1$,  vanishing in the large volume limit.

 The similar
  transformation (\ref{hh-1nt}) of the linear differential form
    \bea\label{chh-2P}
 \int d^3x \sum_{F} P_{\widetilde{F}}\partial_0\widetilde{F}&=&\frac{i}{2}\sum_{F,\textbf{l}}
  \left(F^+_{-\textbf{l}}
  \partial_0F^-_{\textbf{l}}-F^-_{\textbf{l}}\partial_0F^+_{-\textbf{l}}\right)
 +
  \frac{i}{2}\sum_{F,\textbf{l}}
  \left(F^+_{-\textbf{l}}
  F^+_{\textbf{l}}-F^-_{\textbf{l}}F^-_{-\textbf{l}}\right)\partial_0\widetilde{\triangle}_F
\eea
 in action (\ref{4-s}) is not canonical. Therefore, the transition from  field variables
 % to  QFT energy (\ref{chh-2e}) in terms
to the observable quantities (conformal occupation number and one-particle energy)
has physical consequences in the linear form  (\ref{chh-2P}). They are sources
   of creation of pairs from the stable vacuum:
     \bea\label{cF-1}
 \widetilde{\triangle}_{F=v^{T},f}&=&\log\sqrt{\omega_F},~~~~~~~~~\widetilde{\triangle}_{F=v^{||}}=\log \frac{a}{\sqrt{\omega_F}};\\\label{cF-2}
  \widetilde{\triangle}_{F=h,q}&=&\log {a}{\sqrt{\omega_F}},~~~~~~~~~\widetilde{\triangle}_{F=Q,h^{TT}}=\log a;
 \eea
 here
 $v=v^{||}+v^{T}$ are fields of $W$ and $Z$ vector bosons, $f$ are fermions, $h^{TT}$ is graviton, $h$ is a massive
 scalar (Higgs) particle (see  the massive vector theory in detail in \cite{Blaschke_04,hpp}).

The equation (\ref{eta-1}) shows us that the conformal fermion source (\ref{cF-1}) $\log\sqrt{\omega_f}$
differs from the standard one by the term $(3/2)\log a$ which
can lead in SC to intensive creation of massless fermions
 forbidden by observational data and general theorem
 of field theory \cite{JINR-1}.

 In comparison with the classical field theory
with arbitrary occupation numbers considered before,
the new element of QFT is the stable vacuum $b^-_{F,\textbf{l}}|0>=0$,
where $b^-_{F,\textbf{l}}$ is the operator of annihilation of a quasi-particle defined by
 the  Bogoliubov transformation of the operator of particle
 $F^+_{\textbf{l}}=\alpha b^+_{F,\textbf{l}}+\beta^* b^-_{F,\textbf{l}}$,
 so that the equations of motion of the  Bogoliubov quasi-particle become diagonal
 $\partial_\eta b^{\pm}_{F,\textbf{l}}=\pm \omega_b b^{\pm}_{F,\textbf{l}}$, where
 $\omega_b$ is the quasi-particle energy \cite{grib80,ps1}.

  According to these formulae~(\ref{chh-2P}) --
  (\ref{cF-2})
  massless particles, photons and neutrinos, cannot be created in homogeneous  Universe~(see \cite{grib80}).
 There is an estimate in \cite{grib80}
 that fermions and transverse vector bosons  (\ref{cF-1}) are not sufficient, in order
 describe the present-day content of the Universe.
  The creation of gravitons is suppressed by
  the isotropization processes discussed in \cite{grib80}.
 It was shown \cite{Blaschke_04} that just the longitudinal $W$, $Z$ vector bosons are the
  best candidates in SM to form the radiation $(\Omega_{\mathrm{radiation}})$ and
  the baryon matter $(\Omega_{\mathrm{b}})$
  contributions to the Universe energy budget in
the Conformal Cosmological model. The Higgs particle creation is similar to the one of
the longitudinal components of the vector bosons (compare (\ref{cF-2}) and (\ref{cF-1})).

The creation of vector bosons  started at the moment,
 when their wavelength coincided with the horizon length
$  M^{-1}_{\rm v}=(a_{\rm v} M_{0\rm W})^{-1}=
H_{\rm v}^{-1}=a^2_{\rm v} (H_{0})^{-1}$.
This follows from the uncertainty  principle that gives
  the instance of creation of  primordial particles
  \be\label{cr-3}
 a_{\rm v}^3= \frac{H_{0}}{M_{0\rm W}}\simeq 27 \cdot 10^{-45}=(3\cdot 10^{-15})^3~~~~~
 \to ~~~~~ a_{\rm v}\simeq  3\cdot 10^{-15}=(1+z_{\rm v})^{-1}.
 \ee
As it was shown in \cite{Smolyansky_02} using the scalar field model
that taking into account interactions
  \be\label{cr-5h}
   \partial_\eta v^{\pm}(\textbf{k},\eta)= \pm i\omega_vv^{\pm}(\textbf{k},\eta)
    +\partial_\eta \triangle_{v^{|\!|}}(\eta)
   v^{\pm}(\textbf{k},\eta)+i [H_{\rm int}, v^{\pm}(\textbf{k},\eta)]\ee
 can  lead to the collision integral and the Boltzmann-type distribution.
 As a model of such a statistical system,
 a degenerate Bose-Einstein gas was considered in \cite{Blaschke_04}, whose distribution
 function has the form
 $\left\{\exp\left[\frac{\omega_{\rm v}(\eta)- M_{\rm v}(\eta)}{ k_{\rm B}T_{\rm v}}\right]
 -1\right\}^{-1}$
 where $T_{\rm v}$ is the boson temperature treated as
 the measurable parameter of the particle distribution function in the kinetic
 equation with the collision integral.

The value of  the vector boson  temperature
directly follows  from the  analysis of the
numerical calculations in \cite{Blaschke_04}, from the dominance of longitudinal vector
bosons with high momenta  $n(T_{\rm v})\sim T^3_{\rm v}$ and from the fact that the relaxation time
 \cite{bernstein} $ \eta_{\mbox{\small  rel}} =
 \left[{n(T_{\rm v})\sigma_{\mbox{\small scatt}}}\right]^{-1} $
 is equal to the inverse Hubble parameter, if initial data (\ref{cr-3}) is chosen.
In the case of relativistic bosons $n(T_{\rm v})\sim T^3_{\rm v}$ and
   $\sigma_{\mbox{\small scatt}}\sim 1/M_{\rm v}^2$ the vector boson  temperature value
 $ T_{\rm v}\sim (M_{\rm v}^2H_{\rm v})^{1/3}=(M_{0\rm W}^2H_0)^{1/3}\sim 3$~K,
     is close to the observed temperature of the
cosmic microwave background radiation. So the temperature arises in this case
after creation of particles and it is described in the usual way \cite{Smolyansky_02}.
Note that the
masses of those particles is provided by the standard mechanism of
the absorbtion of the extra Higgs field components. The latter happens due to the
nonzero Higgs field vacuum expectation value, which already existed
at the initial moment $\eta=0$, when there were no any particles and hence
no temperature.

In this way CMB inherits the primordial vector boson temperature and density,
 $\Omega _{\rm rad}\simeq M_W^2 \cdot a_I^{-2}= 10^{-34}10^{29} \sim 10^{-5}$.
 In the early epoch with the dominant abundance of weak bosons
(due to the Bogoliubov condensation),
 their Bell-Jackiw-Adler triangle anomaly and the SM CKM mixing in the environment
of the Universe evolution lead to the non-conservation of the sum of lepton and
baryon numbers and to the
 baryon-antibaryon asymmetry of matter in the Universe $\dfrac{n_b}{n_\gamma}\sim X_{\rm CP}\sim
 10^{-9}$.
  The present-day baryon density is calculated by the evolution of the baryon density
  from the early stage, when it was directly related to the photon density.
So that its present--day  value is equal to
  $\Omega _{\rm b}\simeq 10^{-34}10^{-9}  10^{43} (a_{\rm v}/a_L)^3 \simeq  \alpha_W =\alpha_{\mathrm QED}/\sin^2\theta_{W}
 \simeq 0.03$, where the factor $(a_{\rm v}/a_L)^3\simeq \alpha_W$ arise as a
 retardation caused by the life-time of the W-boson~\cite{Blaschke_04}.

 Thus we gave a set of argument in favor of that the GR and SM accompanied by a scalar field $Q$ can describe cosmological creation of the Universe with its matter content  %vacuum occupation numbers
% $<0|F^+_{\textbf{l}}F^-_{\textbf{l}}|0>=|\beta|^2$
% estimated in  GR\&SM are
\bea \label{free-1}
<0|\hat {\textsf{C}}_{\rm U}[P,\widetilde{F}]|0>&=&4V_0^2a^2[a'^2-\rho (a)]=0,
 \eea
in agreement with
 the observational data in (\ref{free-c1}), where
 $\Omega_{\mathrm{rigid}}=0.85\pm 0.10$, $\Omega_{\mathrm{radiation}}\simeq 4 \cdot 10^{-5}$, and
$\Omega_{\mathrm{b}}\simeq 3 \cdot 10^{-2}$,
 if observables (one-particle energy, occupation number, temperature, distance,
 time, {\it etc.}) are identified with conformal variables with inertial initial data \cite{Blaschke_04,zakhy}.
 $\Omega_{\mathrm{CDM}}\simeq 0.3$ can be considered as the input parameter for
fitting $Q$-particle potential parameters.
%Cosmology
 %in the Friedmann equation
% if the initial data (\ref{id-4}) -- (\ref{id-7}) and the conformal variables of CC
 %are chosen.
%So in CC one gets the following values for the particular contributions to the density:

In order to pose the problem of a more accurate calculation that
  can be done in this model in future, one needs to establish
 the parameters of the Higgs potential. It is the topic of the next section.

\section{Higgs field contribution to energy density} \label{sect5}

 The nonzero average of the Higgs field given by the initial conditions
 provides the electroweak symmetry breaking required by SM.
In the Standard Model embedded in the cosmology equations~(\ref{L-1}),
the values of the initial data~(\ref{in-d2})
are directly defined by the other parameters of the model: $\phi_I = {M_W}/{g}_W$.

On the classical level
 the introduction of the initial data for the Higgs field
 allows us  to consider the situation, when
the parameter $c_0\equiv\langle\phi\rangle$ in the  Higgs
potential, so that
\bea \label{higgs-2}
\textsf{V}_{\rm Higgs}(\phi) = {\lambda} \left[\phi^2-\langle\phi\rangle^2\right]^2, \qquad
\textsf{V}_{\rm Higgs}(\langle\phi\rangle)\equiv 0.
\eea
In this way we can remove contribution of the Higgs field zeroth harmonic into
the energy density together with possible creation of monopoles and tachions.

In the perturbation theory loop diagrams lead to the
Coleman--Weinberg potential~\cite{coleman79}, which can
substantially modify the initial {\em classic} potential leading to
the fine tuning problem in the Standard Model.
Contrary to the case of the SM, loop corrections can not shift the position of
the minimum in~(\ref{higgs-2}) because of the symmetry in the potential.

In our case the condition
 \bea \label{eff-1}
\textsf{V}_{\rm eff}(\langle\phi\rangle)= 0
 \eea
is the natural constraint of the unit
vacuum-vacuum transition amplitude at the point of the potential
extremum:
 \bea
 \textsf{V}_{\rm eff}(\langle\phi\rangle)=
 -i {\mathrm Tr}\log\left(<0|0>[\langle\phi\rangle]\right), \qquad
 <0|0>[\langle\phi\rangle] =1 \Longrightarrow \textsf{V}_{\rm eff} (\langle\phi\rangle)=0.
 \eea
In other words, the condition is motivated by the principle of minimization of the vacuum energy
and by the very definition of the classical potential.

So we should have the zero value of the Coleman--Weinberg potential and of its
derivative for $\phi=\langle\phi\rangle$. These conditions
correspond to the vacuum stability boundary in the Standard Model as
discussed in Ref.~\cite{Ford:1992mv}. The boundary has been
extensively studied in the literature (see
review~\cite{Djouadi:2005gi} and references therein). The
corresponding equation can be resolved with respect to the Higgs
mass, which than depends on the masses of top-quark, $Z$ and $W$
bosons, on the EW coupling constants, and on the value of the
cut-off parameter $\Lambda$, which regularizes divergent loop
integrals. The modern
studies~\cite{Altarelli:1994rb,Casas:1994qy,Einhorn:2007rv} which
include complete one-loop with a certain resummation for the running
masses and coupling constants and the dominant two-loop EW
contributions. They are in a reasonable agreement with each other
and give in SM  the following range of the lower Higgs mass limit.
For $\Lambda = 1$~TeV, the improved lower bound
reads~\cite{Einhorn:2007rv}:
 \bea
 m_h[\mathrm{GeV}] > m_h^{\mathrm{bound.}}=
52 + 0.64(m_t(\mathrm{GeV})-175) - 0.5\frac{\alpha_s(M_Z) - 0.118}{0.006}.
 \eea
For very high values of the cut-off $\Lambda \to
10^{19}$~GeV one gets $m_h>m_h^{\mathrm{bound.}}\approx 134$~GeV.
These values $m_h^{\mathrm{bound.}}$
in the Standard Model correspond to the limiting case, where the
model breaks down. On the contrary, in our case these values are
just our predictions for $m_h$:
 \bea \label{range}
 52\ \mathrm{GeV}\ \lsim m_h \lsim 134\ \mathrm{GeV}.
 \eea
Numerous experimental data indirectly support the existence of a
SM-like Higgs particle of a relatively low mass~\cite{:2005ema}:
 $m_h (\mathrm{SM\ fit}) = 129^{+ 74}_{-49} {\ \mathrm{GeV}}$
with the direct experimental limit
 $\label{exp_limit} m_h>114.4$~GeV at the 95\% CL~\cite{pdg}.

So one can see that the Standard Model deserves
{\em new physics} contributions parameterized by the cut-off not lower
than at a rather high energy scale $\sim 100$~TeV.

The domain of Higgs masses below 134~GeV (and higher) will be
studied soon experimentally at the Large Hardon Collider (LHC). Higgs bosons
with such masses decay mainly into pairs of
$b$-quarks~\cite{hhg,Djouadi:2005gi}. As concerns the production
mechanism, for the given range of $m_h$ the sub-process with gluon-gluon
fusion dominates~\cite{Hahn:2006my} and the
corresponding cross sections provide a good possibility to discover
the Higgs boson at the high-luminosity LHC machine.

Real Higgs particles created in the Early Universe were important for
the energy budget of the Universe as described above. The present-day
contribution of  Higgs particles is vanishing, since the production rate described
by Eq. (\ref{cr-5h}) is suppressed for the present-day
value of the Hubble parameter.

The initial data scenario removes the infinite  potential
 vacuum energy density,  creation of monopoles, and
tachion behavior at high energies, because the Higgs potential has form~(\ref{higgs-2})
which can be cast as
 \bea
 {\bf V}_{\rm Higss} \left(\phi=\langle\phi\rangle+\frac{h}{\sqrt{2}}\right)
 = m_h^2 \frac{h^2}{2} + \sqrt{\frac{\lambda}{2}}\, m_h h^3 + \lambda \frac{h^4}{4},\qquad
 \lambda=\frac{g^2_W}{4}\frac{m_h^2}{M_W^2}\sim 0.2 \div 0.3\, .
 \eea

\section{Conclusion} \label{sect6}

The Higgs effect was studied in
the cosmological model  following from
the  emerging  GR\&SM action (\ref{4-s}) supplemented by the additional $Q$ field
under the assumption of the potential-free (inertial) zeroth mode dynamics of both
scalar fields $\textsf{V}_{\rm Higgs}(\langle \phi \rangle)=0,\textsf{V}_{\rm U}(\langle Q\rangle)=0$.
 So that the   potential
vacuum energy density,  possible creation of monopoles, a
tachion behavior at high energies are excluded from the very beginning.
The spontaneous symmetry breaking can be provided by
initial data of the zeroth harmonic of the scalar Higgs field $\langle\phi\rangle=M_W/g_W, \langle\phi\rangle'=0$ without its contribution to the energy density.
The latter can be formed by an inertial motion of the
zeroth harmonic of an additional scalar field $P_{\langle Q\rangle} =2V_0H_0 \sqrt{\Omega_{\rm rigid}}$.
 In the neighborhood of
  the point of cosmological singularity, this motion corresponds to the most singular primary energetic regime of the rigid state. The research of the
 constraint-shell dynamics in terms of the conformal variables shows us that
 at the  point of cosmological singularity there is no any physical sources of the inflation mechanism.

In the limit of cosmological singularity $a=0$, GR and SM
 contain
  the process of vacuum particle creation.
This vacuum particle creation is described as the Bogoliubov vacuum expectation value  of the energy constraint operator. The  estimation of this vacuum expectation value is
in agreement with the observational data, if
observable quantities are identified with the conformal variables \cite{Blaschke_04}.
These variables are distinguished by both the observational Cosmology
and particle creation tool. This Conformal Cosmology
is not excluded by modern observational data including
    chemical evolution and SN data~\cite{Riess:2001gk,Tegmark:2001zc}, if at all these
    epochs the primordial rigid state dominates $\sqrt{\Omega_{\rm rigid}}\sim 1$.

 In the new Inertial scenario the CMB conformal temperature
 is predicted by
 the collision integral kinetic equation  of  longitudinal vector bosons $W$ and $Z$
together with the Higgs particles. The  temperature arises as the consequence of
the primordial particle collisions
 after their creations in the cold Universe filled in by the $Q$ zeroth  harmonic energy density.

In order to pose a problem of more accurate calculation that
  can be done in this model in future, we established
 the parameters of the Higgs potential that follow from the LEP/SLC experimental data.
The present fit of the
LEP/SLC experimental data indirectly supports
 rather low values of the Higgs mass, $114 < m_h \lsim 134$~GeV,
predicted in our approach.

\section{Acknowledgements}

We are  grateful to B.M.~Barbashov, K.A.~Bronnikov,
D.I.~Kazakov, M.Yu.~Khlopov, E.A.~Kuraev, R.~Lednick\'{y}, L.N.~Lipatov,
V.F.~Mukhanov, I.A.~Tkachev, G.~t'Hooft,  and A.F.~Zakharov
for interest, criticism and creative discussions.
One of us (A.A.) thanks for support the grant of the President RF
(Scientific Schools 5332.2006) and the INTAS grant 05-1000008-8328.
Two of us (L.G,V.P.) thank the Bogoliubov-Infeld grant.

\end{document}